\documentclass[reprint,aps,superscriptaddress]{revtex4-2}
\usepackage{graphicx}
\usepackage{tcolorbox}
\usepackage{bm}
\usepackage{xr}
\usepackage{float}
\usepackage{amsmath}
\usepackage{booktabs}
\begin{document}

\title{Detecting informative higher-order interactions in statistically validated hypergraphs}

\author{Federico Musciotto}
\affiliation{Dipartimento di Fisica e Chimica Emilio Segr\`e, Universit\`a di Palermo, Viale delle Scienze, Ed. 18, I-90128, Palermo, Italy}
\author{Federico Battiston}
\affiliation{Department of Network and Data Science, Central European University,  1100 Vienna,. Austria}
\author{Rosario N. Mantegna}
\affiliation{Dipartimento di Fisica e Chimica Emilio Segr\`e, Universit\`a di Palermo, Viale delle Scienze, Ed. 18, I-90128, Palermo, Italy}
\affiliation{Complexity Science Hub Vienna, Josefst\"adter Strasse 39, 1080, Vienna, Austria}

\begin{abstract}
Recent empirical evidence has shown that in many real-world systems, successfully represented as networks, interactions are not limited to dyads, but often involve three or more agents at a time. These data are better described by hypergraphs, where hyperlinks encode higher-order interactions among a group of nodes. In spite of the large number of works on networks, highlighting informative hyperlinks in hypergraphs obtained from real world data is still an open problem. Here we propose an analytic approach to filter hypergraphs by identifying those hyperlinks that are over-expressed with respect to a random null hypothesis, and represent the most relevant higher-order connections. We apply our method to a class of synthetic benchmarks and to several datasets. For all cases, the method highlights hyperlinks that are more informative than those extracted with pairwise approaches. Our method provides a first way to obtain statistically validated hypergraphs, separating informative connections from redundant and noisy ones.
\end{abstract}

\maketitle

\section*{Introduction}
Over the last years, advances in technology have made available a deluge of new data on biological and socio-technical systems, which have helped scientists build more efficient and precise data-informed models of the world we live in. Most of these data sources, from online social networks to the world trade web and the human brain have been fruitfully represented as graphs, where nodes describe the units of the systems, and links encode their pairwise interactions~\cite{boccaletti2006complex}. Yet, prompted by new empirical evidence, it is now clear that in most real-world systems interactions are not limited to pairs, but often involve three or more agents at the same time~\cite{battiston2020networks}. In our brain, neurons communicate through complex signals involving multiple partners at the same time~\cite{petri2014homological,giusti2016two}. In nature, species co-exist and compete following an intricate web of relationships which can not be understood by considering pairwise interactions only~\cite{grilli2017higher}. In science, most advances are achieved by combining the expertise of multiple individuals in the same team~\cite{patania2017shape}.

To fully keep into account the higher-order organization of real networks, new mathematical frameworks have been proposed, rapidly becoming widespread in the last few years. Computational techniques from algebraic topology have made possible to extract the ``shape'' of the data, investigating the topological features associated to the existence of higher-order interactions from social networks to the brain~\cite{patania2017topological,sizemore2019importance}. In parallel, traditional network measures have been generalised to account for the existence of non-pairwise interactions. This includes new proposals for centrality measures~\cite{estrada2006subgraph, benson2019three}, community structure~\cite{carletti2020random}, and simplicial closure, that is a generalisation of clustering coefficient to higher-order interactions~\cite{benson2018simplicial}. The temporal evolution of higher-order social networks has been investigated, showing the presence of non-trivial correlations and burstiness at all orders of interactions~\cite{cencetti2021temporal}. Besides, explicitly considering the higher-order structure of real-world systems has led to the discovery of new collective phenomena and dynamical behavior, from social contagion~\cite{iacopini2019simplicial, neuhauser2020opinion} and human cooperation~\cite{alvarez2021evolutionary} to models of diffusion~\cite{schaub2020random,carletti2020random} and synchronization~\cite{skardal2020higher, millan2020explosive, lucas2020multiorder,gambuzza2021stability}.

Among the several frameworks, hypergraphs, collections of nodes and hyperlinks encoding interactions among any number of units, represent the most natural generalisation of traditional networked structures to explicitly consider systems beyond pairwise interactions~\cite{battiston2020networks, berge1973graphs}. However, mapping data and mathematical frameworks presents us with some new challenges. For instance, for some systems higher-order interactions might be difficult to observe, or only be recorded as a collection of pairwise data. To overcome this limitation, recent work has developed a bayesian framework to reconstruct higher-order connections from simple pairwise interactions following a principle of parsimony~\cite{young2020hypergraph}.

In spite of the explosion of new methods to analyse systems interacting at higher orders, a filtering technique working for hypergraphs is not yet available. Filtering techniques are a relatively recent addition to network analysis. Extracting the filtered elements of a network allows to focus on relevant connections that are highly representative of the system, discarding all the redundant and/or noisy information carried by those nodes and connections that can be described by an appropriate statistical null hypothesis (for example the configuration model of the system). Different names have been proposed so far to address this approach. The first name used was {\it {backbone}} of a network~\cite{serrano2009extracting}. In this case the stress was on the links of nodes that were not compatible with a null hypothesis of equally distributed strength. Another proposal was {\it {statistically validated network}}~\cite{tumminello2011statistically}. A statistically validated network is a subgraph of an original graph where the selected links are those associated with a pair node activity which is not compatible with the one estimated under a random null hypothesis taking into account the heterogeneity of activity of nodes. Statistically based filtering of real networks has been investigated in studies focusing on classic examples of networks such as airports~\cite{serrano2009extracting} and actor/movies~\cite{tumminello2011statistically} networks, trading decisions of investors~\cite{tumminello2012identification,musciotto2016patterns,musciotto2018long,challet2018statistically},  mobile phone calls of large set of users~\cite{li2014statistically,li2014comparative}, financial credit transactions occurring in an Interbank market~\cite{hatzopoulos2015quantifying}, intraday lead-lag relationships of returns of financial assets traded in major financial markets~\cite{curme2015emergence}, the Japanese credit market~\cite{marotta2016backbone}, international trade networks~\cite{straka2017grand}, social networks of news consumption~\cite{becatti2019extracting} and rating networks of e-commerce platforms~\cite{becatti2019entropy}. The procedure of filtering nodes and links in a real network is not unique both in terms of methodology and in terms of null hypothesis. Examples of different approaches have been recently proposed in the literature~\cite{gemmetto2017irreducible,coscia2017network,saracco2017inferring,kobayashi2019structured,marcaccioli2019polya}. Several of these techniques have been reviewed and discussed in~\cite{iori2018empirical,straka2018ecology,micciche2019primer,cimini2019statistical}. 

All works available so far have performed network filtering at the level of pair of nodes. In this work we introduce a new methodology filtering complex systems with interactions of various possible orders. Our approach explicitly takes into account the heterogeneity of the system and therefore it is able to highlight over-expression of hyperlinks of different size and weight. In particular, by mapping each layer of hyperlinks of a specific size in a bipartite system, our method identifies those hyperlinks that are over-expressed with respect to a random null hypothesis. To show the informativeness of our new filtering method, through a synthetic benchmark we show that our approach detects real hyperlinks with higher sensitivity and accuracy than other traditional filtering techniques. We then apply our method to three different empirical social datasets. We show that the results obtained with our analysis are able to highlight information that is not obtainable neither from the unfiltered hypergraphs nor with a pairwise statistically validated analysis of the same system.

\section*{Results}
\begin{figure}[H]
	\centering
	\includegraphics[width=0.5\textwidth]{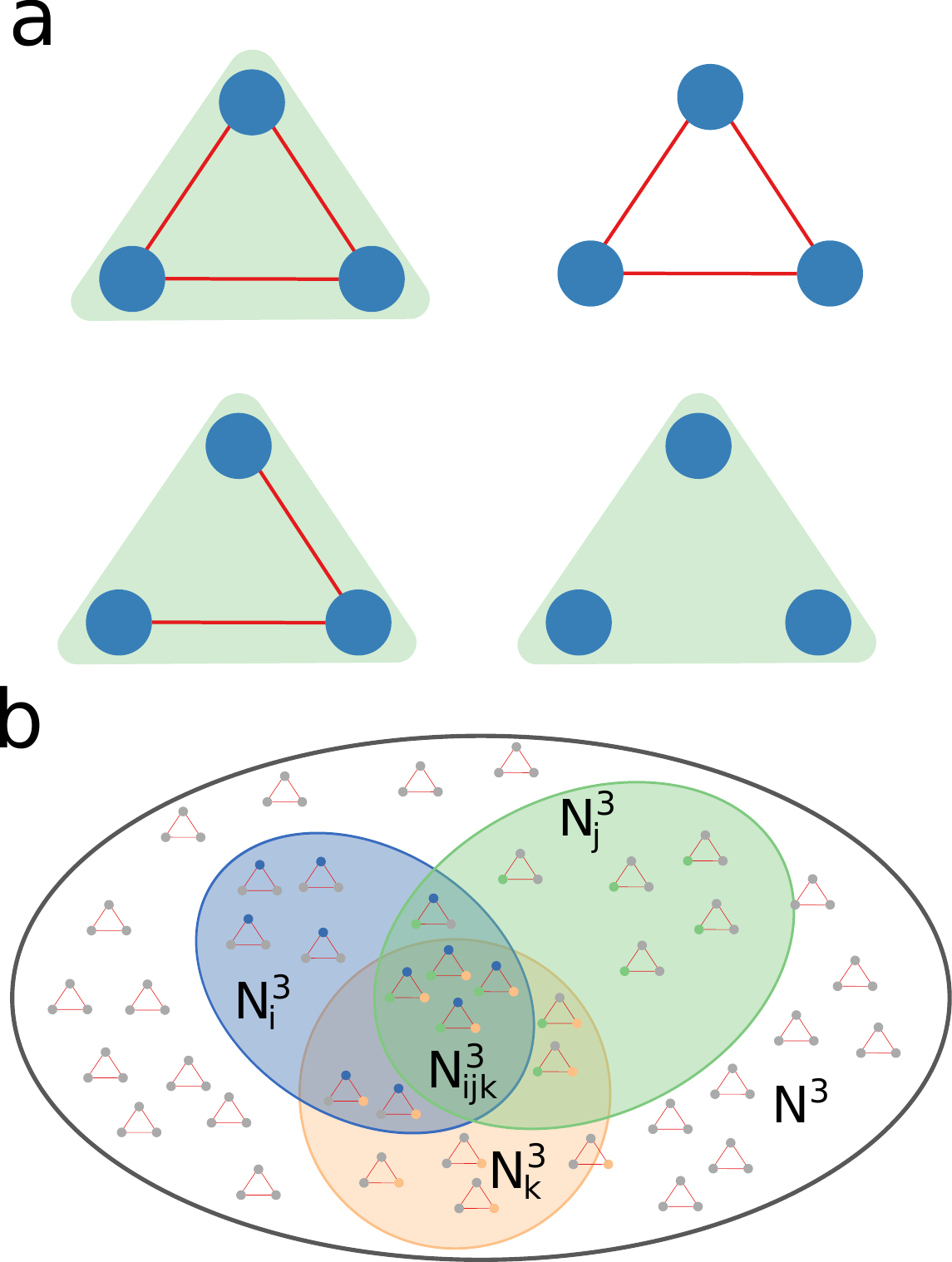}
	\caption{\label{fig:sketch}\textbf{A higher-order filter for hypergraphs. a)} Schematic illustration of false positives and false negatives in the investigation of statistically validated hyperlinks of size $n=3$ detected by SVH compared with an approach based on statistically validated pairwise interactions. Green shaded triangles represent statistically validated hyperlinks with SVH whereas red lines depict statistically validated pairwise interactions. By considering hyperlinks going from left to right and from top to bottom: (i) both the validated hyperlink and the 3-clique of pairwise interactions are validated; (ii) the SVH does not statistically validate the hyperlink whereas a 3-clique among the 3 nodes emerges from the validation of pairwise interactions. Taking the statistically validated pairwise over-expression as an indication of full interaction between nodes provides a conclusion on the over-expression of the hyperlink that is a false positive; (iii) and (iv) the hyperlink is statistically validated but the statistically validated pairwise links do not form a clique (i.e. detecting hyperlinks by evaluating the statistical validation of pairwise interactions would produce a false negative). \textbf{b)} Schematic description of the SVH method for hyperlinks of size $n=3$. In this example, node $i$ has color blue, node $j$ has color green and node $k$ has color orange. The hyperlink $\{i,j,k\}$ is occurring 4 times. By knowing the number of all $N^3$ hyperlinks of size $3$ in the system, and $N_i^3$, $N_j^3$, and $N_k^3$ number of hyperlinks including node $i$, $j$, and $k$ respectively, our methodology allow us to compute the probability of observing $N_{i,j,k}^3$ occurrences (see Eq. (1) of the method session). }
\end{figure}

Traditional network filtering approaches, such as the disparity filter~\cite{serrano2009extracting} or the SVN approach~\cite{tumminello2011statistically}, are not suited for higher-order data, since by design they mistreat all the information on connections beyond pairwise interactions. This implies that cliques of size $n$ highlighted with a pairwise approach might not correspond to genuinely statistically validated hyperlinks, possibly producing both false positive and false negative. Consider, for example, the collaboration network between three authors that have strongly interacted in pairs in their research but have never published a paper altogether. The pairwise analysis might detect a clique of 3 nodes whereas the hyperlink with 3 nodes would not exist, thus generating a false positive. Similarly, false negatives might emerge in the case of an over-expressed hyperlink of size $n$ that is not matched by a clique of validated pairwise links. Figure~\ref{fig:sketch}a illustrates the different possibilities of pairwise validation and hyperlink validation for $n=3$.

Here we propose an analytical filtering technique designed to detect over-expressed hyperlinks of various size.
Our method works with weighted hypergraphs, where groups of nodes are connected through \textit{interactions} (hyperlinks) of any size that are not limited to pairwise links. In particular, we model the weight of an hyperlink composed by $n$ nodes as the intersection of $n$ sets, where each set represents all the hyperlinks in which each node is active (see Figure~\ref{fig:sketch}b for a schematic illustration of the method for $n=3$). 
We are interested in evaluating if the weight of a hyperlink is compatible with a null model in which all nodes are selecting randomly their partners. By using the approach developed in Ref.~\cite{wang2015efficient}, we are able to solve this problem analytically (see Methods), associating a p-value to each hyperlink. Our null model preserves the heterogeneity of degree of all $n$ nodes. This is particularly relevant in the case of hypergraphs whose higher-order degree distribution is strongly heterogeneous. This an ubiquitous characteristic of real systems. Finally, the Statistically Validated Hypergraph (SVH) is obtained by putting together all hyperlinks at different sizes that are validated against our null hypothesis. 

\subsection{Benchmark}

\begin{figure*}[ht!]
\centering
\includegraphics[width=0.95\textwidth]{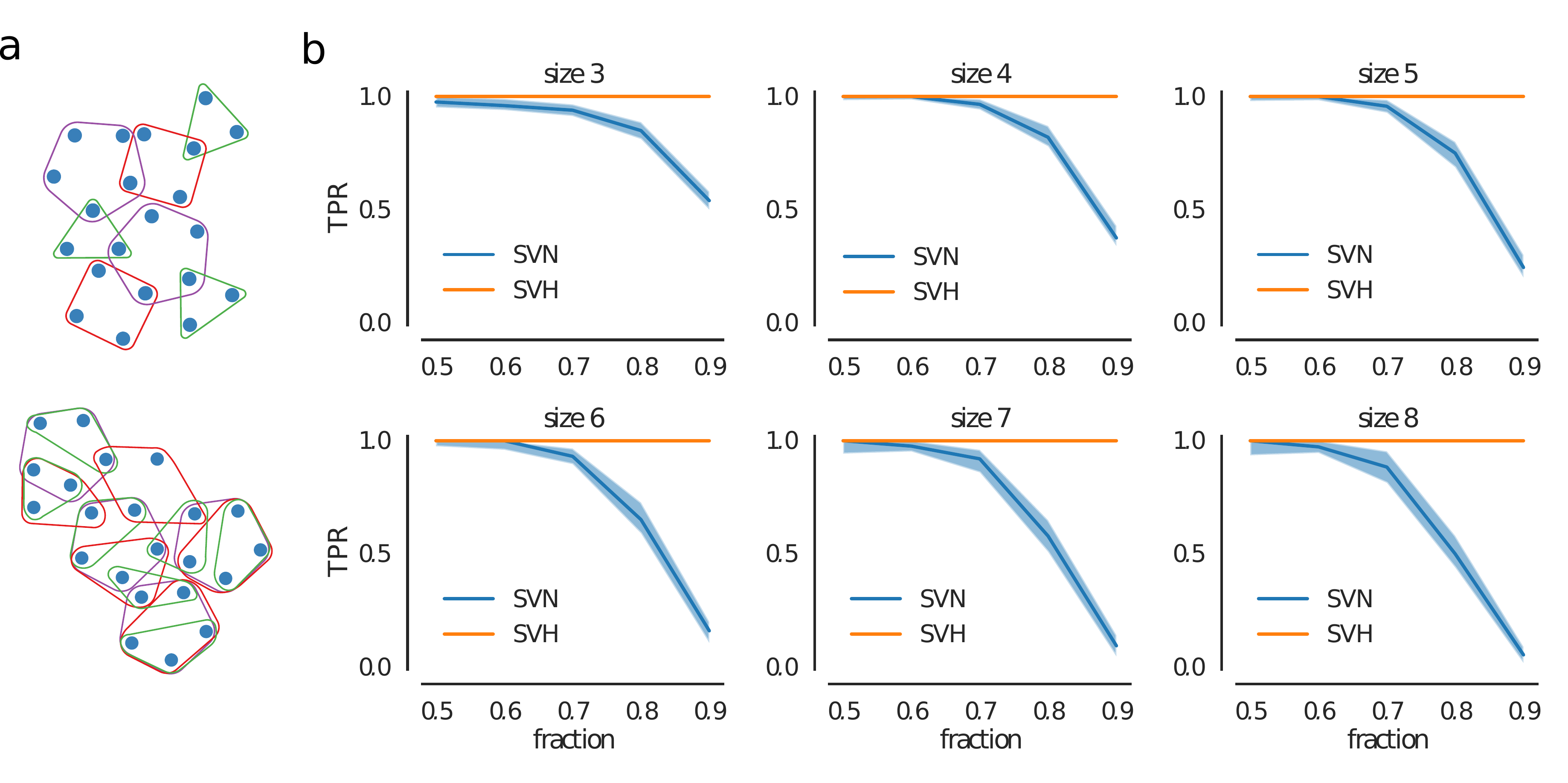}
\caption{\label{benchTPR} \textbf{Hypergraph benchmark and performance analysis. a)} Example of two benchmark realizations with 22 nodes and $f=0.4$ (top) and $f=0.9$ (bottom). The sizes of hyperlinks are 3 (green lines), 4 (red lines) and 5 (purple lines). \textbf{b)} Numerical simulations of the benchmark characterized by the presence of hyperlinks of size $n$ ranging from 3 to 8 (see description in the text). True Positive Rate of the detected hyperlinks obtained by using the SVH  methodology (orange line) and the detection of the pairwise over-expression of SVN (blue line)  as a function of parameter $f$. The shaded area around the lines represents the interval between $10_{th}$ and $90_{th}$ percentile observed in 1000 realizations. Each panel is related to hyperlinks of different size $n$.}
\end{figure*}

We generate hypergraphs that we will use as benchmark in the following way. We select $N$ nodes and a set of sizes for hyperlinks, $\{n\} = \{2,3,...,n\}$. For each size $n$, we select a fraction $f$ of the $N$ nodes, split them in groups of size $n$ and connect each group with $m$ hyperlinks, which in the bipartite representation of hypergraphs correspond to $m$ common neighbors in the set B. Thus, our benchmark is defined by the parameters $(N,m,\{n\},f)$. Here we fix $N=500, m=5, \{n\} = \{2,3,4,5,6,7,8\}$, but the following results are not affected by the specific values of $N$ and $m$ and hold also for larger values of $n$. On the benchmark, we look at the groups of different size that are identified by the SVH (i.e. by our new methodology) and SVN (i.e. inferred by pairwise statistically validated links) approaches when $f$ changes. We choose the SVN as a pairwise approach because it is specifically tailored to work with bipartite networks, that represent the lowest order representation of hypergraphs. In our comparison, for SVH we select the groups of size $n$ defined by validated hyperlinks, whereas for SVN we extract the maximal cliques of size $n$ from the validated projected network. The parameter $f$ affects the probability that a node participates in interactions at different sizes (Figure~\ref{benchTPR}a), thus the higher $f$ the more challenging it is to correctly identify all over-expressed hyperlinks. Indeed, for large $f$ each node is active in groups of different sizes and a filtering method that works only at the pairwise level is likely to produce over-expressed cliques in the SVNs that overestimate the real size of 
an over-expressed  hyperlink because pairwise analysis can merge groups of nodes of different size. 
The true positive rate (TPR) quantifies the fraction of true groups that the two methods are able to identify. To show the performances of the two methods we compute TPR=$\frac{TP}{TP+FN}$ of validated hyperlinks at different sizes as a function of the parameter $f$ (Figure~\ref{benchTPR}b). For each value of $f$ and each size $n$ we generate 1000 realizations of the benchmark and take the median of TPR on all realizations.  While the SVH approach is always able to correctly identify all groups of any size without detecting any false negative no matter the value of $f$, the SVN starts to fail when $f$ grows. Specifically, the rate of false negatives detected by the SVN approach increases with $f$ because SVN tends to merge together groups of different sizes in which the same nodes are active.

To complement this result we also look at the False Discovery Rate (FDR), defined as $FDR = \frac{FP}{FP+TP}$, that quantifies the fraction of false positives on the total number of detected groups (See Figure S1 of the Supplementary Information (SI)). By investigating the benchmark, we find that SVH never detects a False Positive, while the SVN has different performances depending on the value of $f$. 
The behavior observed with the pairwise SVN approach worsens its FDR when $f$ increases because SVN detects groups that are obtained through the aggregation of over-expressed pairwise links even if they are not directly interacting  at the considered size. 

Our benchmark can be perturbed by adding some noise, that simulates random fluctuations or errors in the collection of the data. Specifically, in our simulations we include an additional parameter $p_n$ that represent the probability that each of the $m$ interactions that define a group is randomly assigned to another group of the same size. We study the performance of the two methods as a function of the two parameters and we report the results observed in Figures S2-S3 of the SI. The matrices in the Figures represent the median of the difference in TPR (Figure S2) and FDR (Figure S3) between SVH and SVN as a function of both $f$ and $p_n$. The first column of each matrix represents the difference of the curves plotted in Figures~\ref{benchTPR} and S1. Although the introduction of noise affects the performance of both methods, for all values of $f$ and of the noise parameter $p_n$ the SVH always performs better than the estimation of hyperlinks through the detection of pairwise over-expressed relations with SVN. 

\subsection{US Supreme Court}

In this section we apply the SVH methodology to a dataset that records all votes expressed by the justices of the Supreme Court in the US from 1946 to 2019 case by case~\cite{justice2019database}. This dataset has been extensively investigated in political science to understand and try to predict the patterns of justices' decision by looking at their political alignment during the period in which they were active~\cite{segal1989ideological,epstein1994supreme}. Similar research ideas have started to percolate also the complex systems' community, as shown by a recent work that proposes a link prediction model to forecast the evolution of the citation network spanned by cases ruled by the twin European institution, the Court of Justice~\cite{mones2021emergence}. 

We start noting that such a system naturally fits the framework of hypergraphs, with hyperlinks of size $n$ representing groups of $n$ justices that voted in a case in the same way. As the Supreme Court is made of 9 justices, $n$ can vary from 1 to 9 (in the case of unanimous decisions). In the investigated period we observe 38 different justices judging 8915 cases. We find that the most frequent decisions are the unanimous ones ($\sim$2600), while all the other possible grouping of justices are present with at least 1000 entries (See Table S1 and Figure S4a of the SI). Moreover, we find that the median of the number of decisions that a justice has taken in a group of size $n$ increases with the size of the group (Figure S4b of the SI), signaling that justices are more likely to vote as part of a large majority than in a small minority.
This evidence suggests that an approach that does not take into account interactions beyond the pairwise level is suboptimal to identify groups of justices that show over-expression of voting together, since each justice typically voted in groups of different sizes. In fact, this observed behavior is analog to the behavior seen in the benchmark when we set a large value of $f$. Indeed, in this system when we use both SVH and SVN to detect over-expressed groups at different size (Figure S4c), we find that SVN is unable to find groups at smaller sizes because it is impossible to discriminate groups different from the majority and minority when a pairwise analysis is performed. Moreover, justices vote in the same way in a large number of cases (all the unanimous or almost unanimous ones). Even if we remove the unanimous votes, the situation does not change (See  Figure S4d of the SI). Conversely, the SVH detects a much larger number of over-expressed groups at all possible sizes of interaction. A summary of the numbers of validated groups at different values of $n$ is given in Table S1 of the SI. 

\begin{figure*}[ht!]
	 \centering
	 \includegraphics[width=0.9\textwidth]{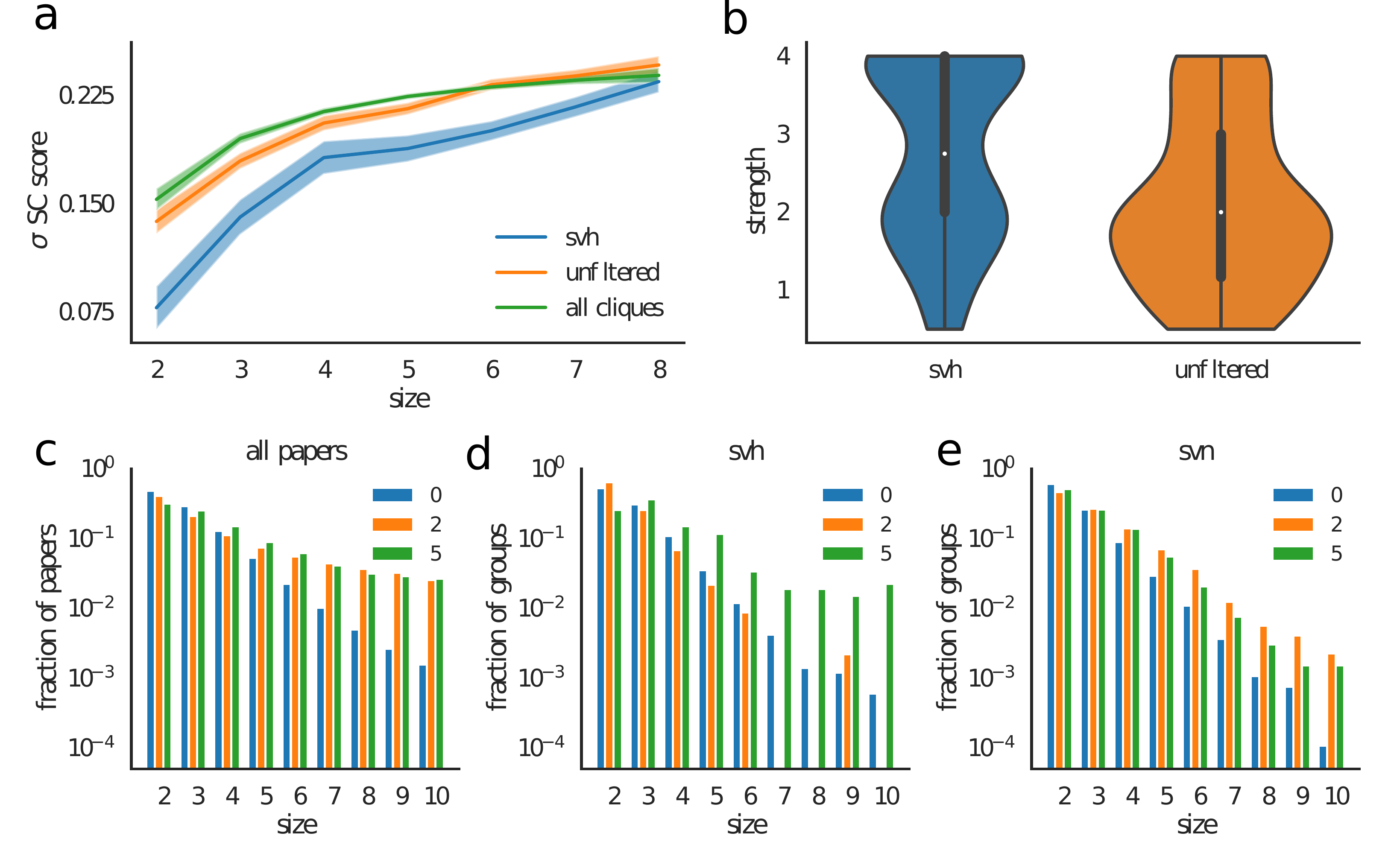}
	 \caption{\label{fig:results}\textbf{Filtering real world hypergraphs. a)} Average standard deviation of the Segal-Cover score as a function of group size of justices of the US Supreme Court (i) for the hyperlinks detected with SVH (blue line), (ii) for the hyperlinks of the unfiltered hypergraph (orange line), and (iii) all possible groups formed by justices that were jointly acting at the Supreme Court during the same time period (green line). The shaded area below the lines represents the standard error of the mean. \textbf{b)} Violin plots of the average strength (as reported in the survey diaries) of (i) the hyperlinks statistically validated by the SVH (blue box) and (ii) the hyperlinks of the unfiltered hypergraph (orange box). \textbf{c)} Fractions of papers written by groups of authors of different size for General Physics (blue bars), Nuclear Physics (orange bars) and Physics of Gases and Plasma PACS categories. \textbf{d),e)} Fractions of over-expressed groups of authors as a function of size $n$ in the over-expressed hyperlinks validated by SVH (\textbf{d}) and of groups of authors that are cliques of size $n$ in the SVN (\textbf{e}) for General Physics (blue bars), Nuclear Physics (orange bars) and Physics of Gases and Plasma (green bars) PACS categories.}
 \end{figure*}

An analysis of groups detected by the SVH provides informative insights on the activity of justices. Indeed, we can characterize each justice with the Segal-Cover (SC) score~\cite{segal1989ideological}, that represents the level of judicial liberalism of each justice throughout her activity in the Supreme Court. 
There is a general SC score and several other scores focusing on specialized categories of legal decisions. When we compute the standard deviation of the SC score for the groups highlighted by the SVH and we compare it with that computed (i) on all the groups of justices observed to vote together at least once and (ii) on all possible groups of justices (to extract the latter we only consider justices that were contemporarily active in the Supreme Court), we find that the groups of justices detected by the SVH have the lowest diversity in liberalism SC score (Figure~\ref{fig:results}a). This means that, with respect to their level of liberalism the groups of justices of size $n$ that present an over-expressed number of joint votes were more similar among them than the set of possible groups of justices. It is worth to note that the SC score is computed exclusively looking at the individual activity of justices case by case, while with the SVH method we are validating groups exclusively looking at their common decisions.

However, the SC liberalism score reduces to a mono-dimensional quantity a piece of information that can be more nuanced. Indeed, the Supreme Court has jurisdiction on cases of different legal areas, ranging from civil rights and criminal procedures to economic decisions. For this reason, the SC score is specialized in a number of distinct scores that capture the attitude of a justice on the different areas~\cite{justice2019database}. Table S2 reports the justices' scores for the three main areas of \textit{criminal procedures, civil rights} and \textit{economic}. As the area of each case is reported in our data, we are able to separately validate groups of justices for the different legal areas. This gives additional insights on the activity of justices. On one side, we find cases as that of justice Antonin Scalia, that was consistently voting with a conservative attitude in all areas, and all the validated groups of size $<5$ in which he is validated are always composed by other conservative justices. On the other side we find more nuanced cases as that of justice Byron White, that was appointed by US president John F. Kennedy. White was progressive on economic issues, and indeed he is present in validated groups of size 3 with two other progressive justices, Thurgood Marshall and William Brennan. Conversely, he had a much more conservative attitude on issues related to civil rights and criminal procedures, and this is detected by our approach: in cases related to these areas he is validated in groups of size 3 with the conservative justices Warren Burger, William Rehnquist and John Marshall Harlan. A summary of the hyperlinks validated in the SVHs of the different areas is reported in Tables S3, S4 and S5.   

\subsection{High School data}

In this section we detect the over-expressed hyperlinks of social interactions observed between students during their stay at a French high school~\cite{mastrandea2015contact}. This data is part of the SocioPatterns project, that aimed at integrating social network analysis traditionally performed on surveys with actual contact data tracked through radio-frequency identification sensors. The contacts are detected and stored when pairs of students   are physically located nearby at a given time $t$. Recording was occuring with a temporal resolution of 20$s$. The data contains also information about self-reported contacts and friendship and Facebook networks that were present among students. These data have already been analysed to understand the overlap between network and contact data~\cite{mastrandea2015contact}. Here we focus on the detection of higher-order interactions from the tracked contacts. In fact, it is important to track the presence of higher-order interactions in a social system as it has an impact on the dynamical processes that can occur on top of it~\cite{iacopini2019simplicial,alvarez2021evolutionary,lucas2020multiorder}. 
In order to build a hypergraph, we extract  the higher-order interactions from the raw data. To do so, for each time step $t$ we build the graph of interactions occurring at time $t$ and we extract the maximal cliques of any size. Indeed, if $n$ students are tracked in a fully connected clique at time $t$, it means that they had a collective interaction at that time.
We find that on top of pairwise interactions the network contains hyperlinks that involve up to 5 students interacting at the same time (Figure S5a). After extracting the over-expressed hyperlinks with both SVH and SVN, we find that, as in the case of US justices, the two methods have a different distribution of validated groups (Figur S5b). Specifically, SVH detects much more groups than SVN at lower size but signicantly less at higher size. We verify that most of the groups detected by SVN at higher size are spurious (Figure S5c), which means that with SVN we validate groups of size $n$ even if the corresponding students were never simultaneously interacting altogether in a group of that size. In this system, the cliques detected by the SVN do not represent a reliable proxy of the over-expressed hyperlinks.

In order to understand the nature of the hyperlinks validated by the SVH, we analyze the validated groups using the available metadata. We use the contact diaries that were filled by the students at the end of a specific day of data collection. This additional information stores contacts that were self reported by students themselves and can be seen as robust information about the system, since it contains interactions that were strong enough to be remembered by the students. We use the diaries to extract cliques at different sizes of interacting students, and we compare this information with the hyperlinks present in the unfiltered dataset and with those validated with the SVH approach. To maintain consistency across the two datasets, we drop contact data that do not contain students that filled the diaries surveys and drop from the diaries contacts that were not tracked by the sensors. Furthermore, we limit contact data to that recorded in the same day of the diaries survey. We find that the SVH is very precise  in retrieving the self reported cliques  (it contains less ''spurious`` hyperlinks that do not correspond to diaries cliques) but it is not highly accurate (i.e. it contains only a smaller fraction of diaries cliques) (Table S6 of the SI).

The reported contacts come with a discrete weight provided by the students themselves that represents the duration of each reported interaction ((i) at most 5 min if $w = 1$, (ii) between 5 and 15 min if $w = 2$, (iii) between 15 min and 1 h if $w = 3$, (iv) more than 1 h if $w = 4$). For each clique in the contact diaries that is also detected in the unfiltered hypergraph or in the SVH we compute the overall strength by averaging the strengths of the links that constitute it. We find that the distribution of the average strength of SVH hyperlinks is higher than the one of unfiltered groups (Figure~\ref{fig:results}b), showing that the SVH hyperlinks detect the most persistent groups. The difference in the distribution of average strength is statistically significant according to a non-parametric Kruskal-Wallis test with score $\sim18$ and $p<0.0001$. It is worth to stress that the hyperlinks present in the SVH do not necessarily correspond exclusively to the interactions with the highest weight. Indeed, we find that hyperlinks of the same weight can be present or absent in the SVH, depending on the heterogenous activity of the involved nodes (Figure S6). In fact, for less active nodes hyperlinks with a small weight are more likely to be validated, while hyperlinks that involve more active nodes need a higher weight in order to be validated.

\subsection{Physics authors}

In this section we analyse the hypergraph of scientific collaborations among Physics authors. To do so, we investigate the APS dataset, that contains authorship data on papers published on journals of the APS group from 1893 to 2015. This dataset has already been extensively investigated to characterize structural and dynamical properties of scientific collaborations, with respect to both authors' careers and topics' evolution~\cite{radicchi2011rescaling, battiston2019taking, chinazzi2019mapping}. 
Here we match the papers present in APS dataset with the Web of Science database using the doi and identify the authors with the ID curated by WoS to maximize disambiguation, which is a well known issue that affects the accuracy of the dataset. Since we are interested in interactions among authors, we limit our investigation to papers with at maximum number of 10 authors to avoid larger collaborations for which direct interaction between all authors are less likely. From the APS dataset we retrieve the PACS of each paper. This allows us to split the set of papers in 10 subfields of physics by using the highest hierarchical level of PACS classification. We focus on the papers published from 1985 onwards as from this year reporting one of more PACS per paper became compulsory. This leaves us with 269,887 papers and 114,856 authors. 

First, we look at the distribution of papers of different size for each subfield (Figure~\ref{fig:results}c). Here we focus on the categories of General Physics (PACS hierarchical integer number 0), Nuclear Physics (PACS number 2) and Physics of Gases and Plasma (PACS number 5) but the CDFs for all PACS categories are shown in Figure S7 of the SI . As expected, we find that PACS have different distributions of team size that highlight different publication habits of the researchers publishing in different PACS categories. In our selection, we find that the subfield with higher percentages of smaller groups is General Physics. On the other side, in Nuclear Physics and Physics of Gases and Plasma there are higher percentages of larger research groups (in Figure~\ref{fig:results}c all percentages sum to 1 because we are cutting from the distributions all papers that are written by groups larger than 10). 

We then apply to the dataset both SVH and SVN methodologies, and extract the distribution of the validated groups with the two methods (Figure~\ref{fig:results}d and Figure~\ref{fig:results}e respectively). At first sight, we find that the distributions of the groups validated with the SVH show relevant differences with the original ones, while with the SVN we obtain similar trends. Specifically, we find that for Nuclear Physics the fraction of over-expressed groups in the SVH goes rapidly to 0 when the size increases, showing an opposite trend if compared to the distribution of the number of authors per paper. This means that the size of most of the SVH validated groups for this PACS is relatively small, in spite of the fact that there are many papers written by larger numbers of authors. In the case of over-expressed groups of the SVN, the distribution for this category of PACS is similar to that of Figure~\ref{fig:results}c. Conversely, PACS 5 (Physics of Gases and Plasma) maintains a similar profile across the different distributions. 

In order to understand this finding we looked, for each size, at the relationship across PACS between the fraction of validated groups and the average number of papers written by a group of that size (Figures S8 for SVH and Figures S9 for SVN). We find that in the case of SVH these two quantities are strongly correlated, with Pearson coefficient ranging from 0.84 to 0.99 and being always statistically significant. This means that with the SVH we validate more groups of authors when these groups are writing on average more papers together. This is a result showing the reliability of SVH results. It is interesting to note that PACS 5 has among the highest average numbers of papers per group at higher sizes, making it clear why the SVH validates more groups at these sizes. On the other hand, in Nuclear Physics most (for some sizes even all) of the detected groups at higher sizes wrote only one paper altogether, so the number of validated groups is much lower. Conversely, with the SVN approach the fraction of detected groups is not related to the average activity of groups of that size (in the scatter plots of Figure S9, all correlations between the fraction of detected groups and teh average number of papers per group are not statistically different from zero). Due to the fact that with the pairwise SVN approach all groups are obtained through aggregation of pairwise over-expressed links, the details of higher-order interactions are missed.

Summing up, the SVH approach gives us an insight that is not evident in the raw data or with methods that are limited to the characterization of pairwise interactions: we find that research areas like Nuclear Physics and Physics of Gases and Plasma are similar with respect to the distribution of papers that are written by research groups of different sizes, but the research groups in the two PACS have different publication habits. In Physics of Gases and Plasma it is more likely that exactly the same group publishes more papers together (and with SVH we identify the groups that do so in a significant way), while in Nuclear Physics most of the medium size collaborations produce a paper just in a single occurrence.

\section*{Discussion}

In the last decade, a deluge of new data on biological and socio-technical systems has become available, showing the importance of filtering techniques able to highlight potentially informative network structures. Recently, hypergraphs have emerged as a fundamental tool to map real-world interacting systems. Yet, extracting the relevant interactions from higher-order data is still an open problem. In this work we proposed Statistically Validated Hypergraphs (SVH) as a method to identify the most meaningful relations between entities of a higher-order system, reducing the complexity carried by noisy and/or spurious interactions.

Our method is able to quantify the probability that an observed hyperlink is compatible with a process in which all involved nodes were randomly selecting their counterparts, reducing the detection of false negatives and false positives occurring with basic pairwise filters. Besides, the null model that we developed naturally reproduces the heterogeneous activity of each node, a crucial feature that overcomes the limitations of a threshold based filtering approach. We have showcased the application of our method to three different systems: the US Supreme Court, the social connections of students interacting in a French high school and the scientific collaborations of Physics authors publishing in journal of the American Physical Society. In all cases, statistically validated groups carry more coherent information than that observed in the unfiltered hypergraphs. For the US Supreme Court, groups of justices with more similar SC profiles are highlighted. For the students of a French high school, groups of students characterized by an intense social interaction are detected, and for the authors of physics papers, the analysis of the SVH unveils a difference in publication habits across subfields that is not evident when looking at the complete system.

A foreseeable development of our methodology is a generalization capable to take into account the temporal dynamics leading to the emergence of a hypergraph, similarly to what was proposed in Ref.~\cite{kobayashi2019structured} for pairwise interactions only. Taken together, we believe that our method, separating meaningful connections from less informative node interactions, is a powerful tool capable to capture the different nuances of higher-order interacting systems.

\begin{widetext}
\section{Methods}
In a SVH, each hyperlink of size $n$ represents a group of $n$ nodes that is over-expressed by comparing its occurrence with that of a null hypothesis that reproduces random group interactions. To extract the p-value of a hyperlink of size $n$, we select a subset of the hypergraph considering only hyperlinks of size $n$, and we compute the weighted degree of each node with respect to this subgraph, $N_{x_1}^n,N_{x_2}^n,...,N_{x_n}^n$. We then extract the weight of the hyperlink connecting all $n$ nodes, $N^n_{x_1...x_n}$ and the total number of hyperlinks of size $n$, $N^n$. We can then assess the probability of $N^n_{x_1...x_n}$ being compatible with a null model where each of the nodes in the group randomly selects its hyperlinks from the whole set of hyperlinks of size $n$. To illustrate the method we start with the simplest case, $n=3$, with three nodes $i$, $j$ and $k$ being active respectively in $N_i^3$, $N_j^3$ and $N_k^3$ interactions of size 3 (Figure~\ref{fig:sketch}b). The probability of having the three nodes interacting together $N^3_{ijk}$ times under a random null model is written as 

\begin{align*}
	p(N^3_{ijk}) &= \sum_{X} H(X|N^3,N_i^3,N_j^3)\times H(N_{ijk}^3|N^3,X,N_k^3) \\
				 &= \frac{1}{\binom{N^3}{N^3_j}\binom{N^3}{N^3_k}}\sum_X \binom{N^3_i}{X}\binom{N^3-N^3_i}{N^3_j-X}\binom{X}{N^3_{ijk}}\binom{N^3-X}{N^3_k - N^3_{ijk}},
\stepcounter{equation}\tag{\theequation}\label{eq:prob3}
\end{align*}
where $H(N_{AB}|N,N_A,N_B)$ is the hypergeometric distribution that computes the probability of having an intersection of size $N_{AB}$ between two sets $A$ and $B$ of size $N_A$ and $N_B$ given $N$ total elements. 

The probability $p(N^3_{ijk})$ in Eq.~\ref{eq:prob3} is obtained through the convolution of two instances of the hypergeometric distribution. Indeed, to compute $p(N^3_{ijk})$ we start from the probability of having an intersection of size $X$ between nodes $i$ and $j$ and multiply it with the probability of having an intersection of size $N_{ijk}^3$ between node $k$ and the intersection set of size $X$ between $i$ and $j$. This product is then summed over all possible values of $X$, i.e. all possible intersections between $i$ and $j$ which are compatible with the observed number of interactions between all the three nodes. Starting from Eq.~\ref{eq:prob3} we then compute a p-value for the hyperlink connecting $i$, $j$ and $k$ through the survival function,
\begin{equation}
\label{eq:pvalue_3}
	p(x\geq N_{ijk}^3) = 1 - \sum_{x=0}^{N_{ijk}^3 - 1} p(x).
\end{equation}
The p-value provides the probability of observing $N_{ijk}^3$ or more occurrences of the hyperlink composed by $i,j,k$. Once all the p-values for all observed hyperlinks are computed, they are tested against a threshold of statistical significance $\alpha$. In all the results presented in this paper we use $\alpha=0.01$. The statistical test is performed by using the control for the False Discovery Rate~\cite{benjamini1995controlling} as a multiple hypothesis test correction. The total number of test considered is $N_t=\binom{N^3_{nodes}}{3}$ which is the number of all possible triplets of the $N^3_{nodes}$ elements that are active in hyperlinks of size 3. Thus, when applying the control for the False Discovery Rate method we start from a Bonferroni threshold computed as $\alpha_B=\alpha/N_t$.

For an hyperlink of generic size $n$, Eq.~\ref{eq:prob3} becomes
\begin{align*}
	p(N^n_{x_1...x_n}) = &\sum_{X_{x_1x_2}} H(X_{x_1x_2}|N^n,N_{x_1}^n,N_{x_2}^n)\times\sum_{X_{x_1x_2x_3}} H(X_{x_1x_2x_3}|N^n,X_{x_1x_2},N_{x_3}^n)\times... \nonumber \\
						 &...\times\sum_{X_{x_1x_2...x_{n-1}}}H(X_{x_1x_2...x_{n-1}}|N^n,X_{x_1x_2...x_{n-2}},N_{x_{n-1}}^n)\times H(N^n_{x_1x_2...x_{n}}|N^n,X_{x_1x_2...x_{n-1}},N_{x_{n}}^n).
\stepcounter{equation}\tag{\theequation}\label{eq:probn}
\end{align*}
As in the case with 3 nodes, the main idea of Eq.~\ref{eq:probn} is to write the overall probability of having an intersection of size $N^n_{x_1x_2...x_n}$ between the activity of $n$ nodes as the product of multiple probabilities of hierarchical pairwise intersections, summed over all the possible configurations compatible with $N^n_{x_1x_2...x_n}$. The specific order of the nodes in the hierarchical intersections does not affect the value of $p(N^n_{x_1...x_n})$. From Eq.~\ref{eq:probn} we extract a p-value as in Eq.~\ref{eq:pvalue_3}. The number of tests to consider when correcting for multiple testing is $N_t=\binom{N_{nodes}^n}{n}$. In our numerical computation we use the approach developed in Ref.~\cite{wang2015efficient}. The approach is analytic but require heavy combinatorial computation, and it might be of difficult application when the hyperlinks are of size larger than about fifteen nodes. The code to use our method is available upon request and will be uploaded on a public repository when the paper is published.

\end{widetext}

\end{document}

\title{Supplementary Information\\ {\large Detecting informative higher-order interactions in statistically validated hypergraphs}}

\author{Federico Musciotto}
\author{Federico Battiston}
\author{Rosario N. Mantegna}

\flushbottom
\maketitle

\section{Benchmarks}
\label{sectionSI:bench}
\begin{figure*}[ht]
\centering
\includegraphics[width=1\textwidth]{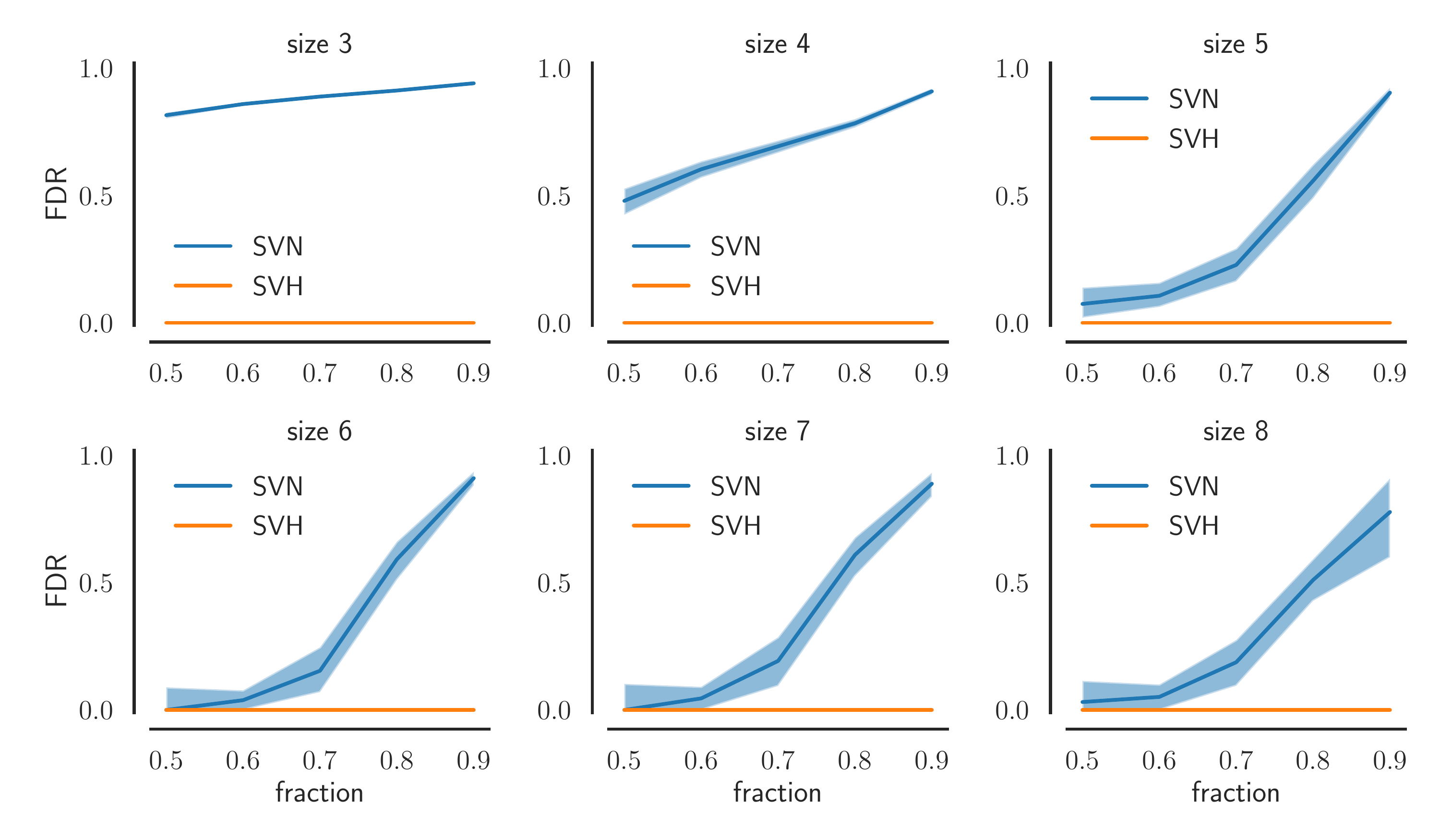}
\caption{\label{fig:FDRbench} Numerical simulations of the benchmark characterized by the presence of hyperlinks of size $n$ ranging from 3 to 8 (see description in the main text). False Discovery Rate of the detected hyperlinks obtained by using the SVH  methodology (orange line) and of the groups inferred by the pairwise over-expression of SVN (blue line)  as a function of parameter $f$. The shaded area around the lines represents the interval between $10_{th}$ and $90_{th}$ percentile observed in 1000 realizations. Each panel is related to hyperlinks of different size $n$.}
\end{figure*}

\begin{figure}[ht]
\centering
\includegraphics[width=1\textwidth]{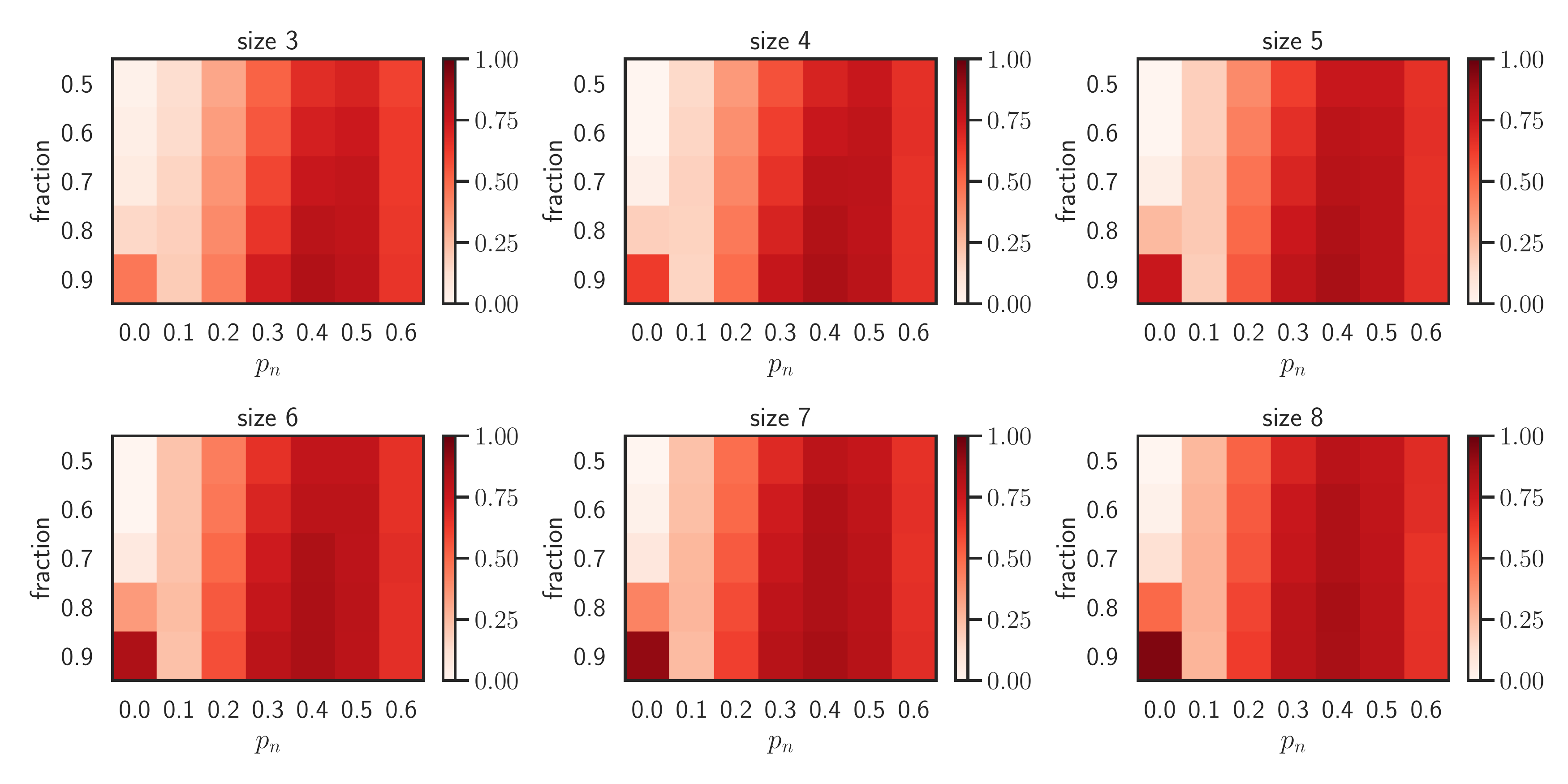}
\caption{\label{fig:TPRmat} Numerical simulations of the benchmark characterized by the presence of hyperlinks of size $n$ ranging from 3 to 8 (see description in the main text) as a function of the fraction $f$ of selected nodes. Median of the difference of TPR of the detected hyperlinks obtained by using the SVH  methodology and the detection of the pairwise over-expression of SVN as a function of parameter $f$ (x-axis) and $p_n$ (y-axis) over 1000 realizations. Each panel is related to hyperlinks of different size $n$.}

\end{figure}

\begin{figure}[ht]
	\centering
	\includegraphics[width=1\textwidth]{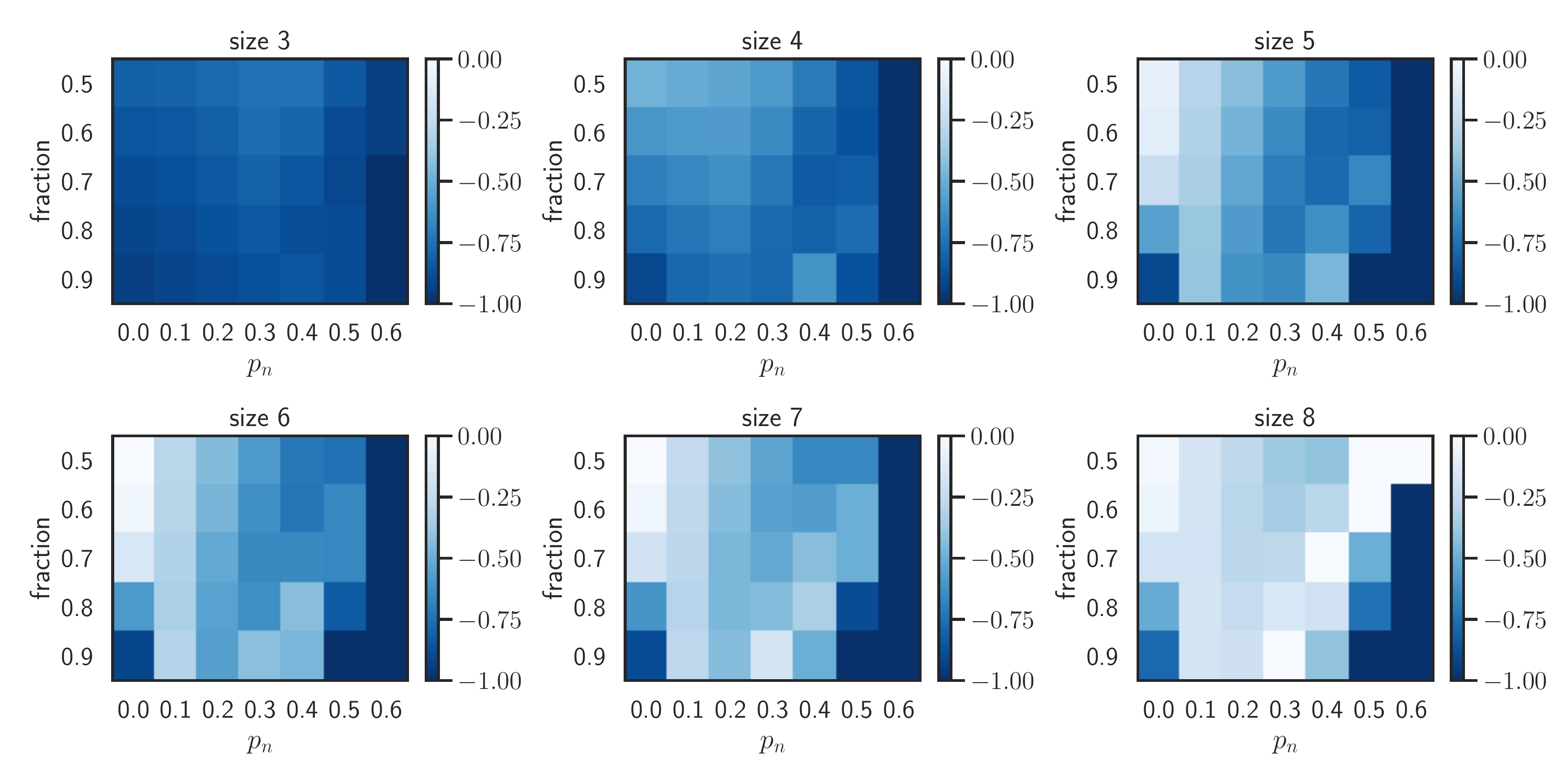}
	\caption{\label{fig:FDRmat} Numerical simulations of the benchmark characterized by the presence of hyperlinks of size $n$ ranging from 3 to 8 (see description in the main text). Median of the difference of FDR of the detected hyperlinks obtained by using the SVH  methodology and the detection of the pairwise over-expression of SVN as a function of parameter $p_n$ (x-axis) and $f$ (y-axis) over 1000 realizations. Each panel is related to hyperlinks of different size $n$.}
\end{figure}

In order to assess the rate of type I errors detected by the SVH on our benchmark we compute the False Discovery Rate (FDR), defined as $FDR = \frac{FP}{FP+TP}$. The behavior of FDR for SVH and SVN for hyperlinks of different sizes is shown in Figure~\ref{fig:FDRbench} over 1000 realizations of the benchmark. After defining the parameter $p_n$ that tunes the insertion of noise in the benchmark, we studied the performance of our method for different values of the parameter. Figure~\ref{fig:TPRmat} and \ref{fig:FDRmat} show the median of the difference in TPR and FDR between SVH and SVN as a function of parameters $f$ and $p_n$.

\clearpage
\section{US Supreme Court}
\begin{figure}[ht]
	\centering
	\includegraphics[width=1\textwidth]{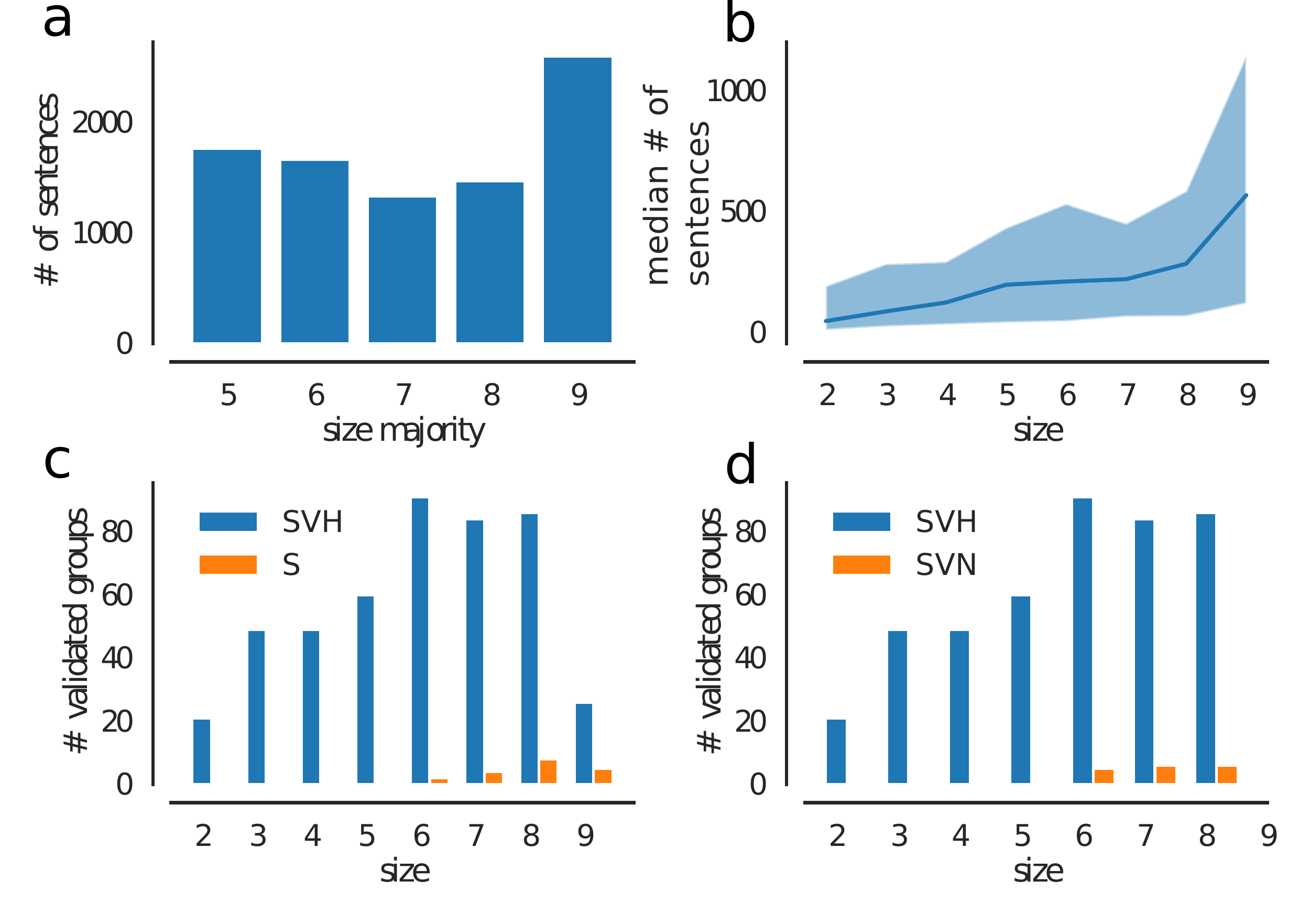}
	\caption{\label{fig:supcourt} \textbf{a)} Number of sentences as a function of the size of majority in the US Supreme Court. \textbf{b)} Median of the number of sentences made by a justice as a function of the size of the co-voting group (blue line). The shaded area around the line represents the interval between $10_{th}$ and $90_{th}$ percentile of the distribution of all justices. \textbf{c)} Number of validated groups in the SVH (blue bars) and in the SVN (orange bars) as a function of group size. \textbf{d)} Same of c) after excluding all unanimous votations when extracting the groups with the SVN approach.}
	
\end{figure}

The US Supreme Court is a system characterized by a high density of interactions, with a relatively small number of justices (38) interacting over 8915 sentences. Table~\ref{tab:justiceSummary} offers a breakdown of the activity of justices by reporting the number of justices and groups active in co-decisions of different sizes, together with the number of hyperlinks validated in the SVH. Other evidences of this high density are given by Figure~\ref{fig:supcourt}a, that plots the number of sentences as a function of majority size and Figure~\ref{fig:supcourt}b, that shows the median number of voting decisions in groups of different sizes made by the justices. This high density results in a poor performance of the SVN method in detecting validated hyperlinks, as shown in Figure~\ref{fig:supcourt}c. Figure~\ref{fig:supcourt}d shows how the performance of SVN remains poor also after dropping all unanimous sentences from the data. 

In Figure 3a of the main text we show how the groups of justices validated by the SVH approach have a higher similarity with respect to the Seagal Cover score. In Table~\ref{tab:justiceScores} we report the SC score of all the 38 justices active in the dataset. Together with the SC score we also report the scores on three separate areas, (i) Criminal procedures, (ii) Civil rights and (iii) Economic issues. A summary of the activity of justices and the outcome of SVH for these three areas is presented in Tables~\ref{tab:justiceScores1},~\ref{tab:justiceScores2} and~\ref{tab:justiceScores8}.
\begin{table}
\small
\centering
\begin{center}
\begin{tabular}{rrrlll}
\toprule
 Size &  Justices &  Votes & Possible groups & Observed groups & Validated groups \\
\midrule
    1 &        34 &   1089 &               - &               - &                - \\
    2 &        37 &   1390 &             264 &             180 &               21 \\
    3 &        38 &   1621 &             868 &             354 &               49 \\
    4 &        38 &   1447 &            1666 &             358 &               49 \\
    5 &        38 &   1764 &            2016 &             465 &               60 \\
    6 &        38 &   1668 &            1568 &             484 &               91 \\
    7 &        38 &   1336 &             764 &             360 &               84 \\
    8 &        38 &   1474 &             213 &             167 &               86 \\
    9 &        38 &   2601 &              26 &              26 &               26 \\
\bottomrule
\end{tabular}
\end{center}
\caption{\label{tab:justiceSummary} Summary of the co-decisions (i.e. same type of vote) of justices in groups of different size. For all sizes we report the number of justices active at least once and the number of votations in which groups of that size were observed. For all sizes larger than 1 we report the number of possible groups of co-voting justices (taking into account as a contrain the period of co-activity of justices), the number of observed groups and the number of groups validated in the SVH.}

\end{table}

\clearpage
\begin{table}
\small
\centering
\begin{center}

\begin{tabular}{lcccc}
\toprule
         name &  SC score & Criminal procedure & Civil rights & Economic \\
\midrule
      HLBlack &     0.875 &              0.668 &        0.701 &    0.811 \\
       SFReed &     0.725 &              0.318 &        0.471 &    0.658 \\
 FFrankfurter &     0.665 &                0.5 &        0.622 &     0.54 \\
    WODouglas &     0.730 &              0.804 &        0.872 &    0.796 \\
      FMurphy &     1.000 &              0.805 &         0.82 &    0.759 \\
    RHJackson &     1.000 &              0.378 &        0.519 &    0.513 \\
   WBRutledge &     1.000 &              0.738 &        0.723 &    0.748 \\
     HHBurton &     0.280 &              0.284 &        0.433 &    0.533 \\
     FMVinson &     0.750 &               0.27 &        0.453 &    0.599 \\
      TCClark &     0.500 &              0.321 &        0.535 &    0.725 \\
      SMinton &     0.720 &              0.231 &        0.362 &     0.62 \\
      EWarren &     0.750 &              0.733 &         0.81 &    0.813 \\
     JHarlan2 &     0.875 &              0.378 &         0.42 &    0.367 \\
    WJBrennan &     1.000 &              0.762 &        0.825 &    0.698 \\
  CEWhittaker &     0.500 &              0.398 &        0.442 &    0.336 \\
     PStewart &     0.750 &              0.449 &        0.484 &     0.45 \\
      BRWhite &     0.500 &              0.329 &        0.552 &    0.593 \\
   AJGoldberg &     0.750 &               0.75 &         0.98 &    0.632 \\
      AFortas &     0.845 &              0.815 &        0.821 &    0.722 \\
    TMarshall &     1.000 &              0.801 &        0.845 &     0.63 \\
     WEBurger &     0.115 &               0.19 &        0.374 &     0.45 \\
   HABlackmun &     0.115 &              0.423 &        0.621 &     0.55 \\
     LFPowell &     0.165 &              0.284 &        0.403 &    0.456 \\
  WHRehnquist &     0.045 &              0.196 &        0.306 &    0.494 \\
    JPStevens &     0.250 &              0.669 &        0.643 &    0.562 \\
    SDOConnor &     0.415 &              0.264 &        0.454 &    0.465 \\
      AScalia &     0.000 &              0.274 &        0.306 &    0.462 \\
    AMKennedy &     0.365 &              0.333 &         0.44 &    0.446 \\
     DHSouter &     0.325 &              0.572 &        0.693 &    0.542 \\
      CThomas &     0.160 &              0.224 &         0.25 &    0.462 \\
   RBGinsburg &     0.680 &              0.623 &          0.7 &    0.566 \\
     SGBreyer &     0.475 &              0.556 &        0.697 &      0.5 \\
    JGRoberts &     0.120 &              0.311 &        0.424 &    0.445 \\
      SAAlito &     0.100 &              0.186 &        0.388 &    0.424 \\
   SSotomayor &     0.780 &              0.669 &        0.707 &      0.5 \\
       EKagan &     0.730 &              0.625 &        0.719 &      0.5 \\
    NMGorsuch &     0.110 &                  0 &          0.5 &    0.333 \\
  BMKavanaugh &     0.070 &                  - &            - &        - \\
\bottomrule
\end{tabular}
\end{center}
\caption{\label{tab:justiceScores} Seagal Cover score of all the 38 justices present in our dataset. The second, third and fourth columns report the scores related to votations concerning respectively (i) criminal procedures, (ii) civil rights and (iii) economic issues.}

\end{table}

\begin{table}
\small
\centering
\begin{center}
\begin{tabular}{rrrlll}
\toprule
 Size &  Justices &  Votes & Possible groups & Observed groups & Validated groups \\
\midrule
    1 &        25 &    202 &               - &               - &                - \\
    2 &        36 &    360 &             264 &              82 &                9 \\
    3 &        37 &    435 &             868 &             141 &               22 \\
    4 &        38 &    433 &            1666 &             130 &               24 \\
    5 &        38 &    505 &            2016 &             180 &               32 \\
    6 &        38 &    436 &            1568 &             180 &               25 \\
    7 &        38 &    321 &             764 &             134 &               23 \\
    8 &        38 &    273 &             213 &              89 &               21 \\
    9 &        38 &    481 &              26 &              25 &               24 \\
\bottomrule
\end{tabular}
\end{center}
\caption{\label{tab:justiceScores1} Summary of the co-decisions of justices in groups of different size for the votations related to criminal procedures. For all sizes we report the number of justices active at least once and the number of votations in which groups of that size were observed. For all sizes larger than 1 we report the number of possible groups of co-voting justices (taking into account as a contrain the period of co-activity of justices), the number of observed groups and the number of groups validated in the SVH.}

\end{table}

\begin{table}
\small
\centering
\begin{center}
\begin{tabular}{rrrlll}
\toprule
 Size &  Justices &  Votes & Possible groups & Observed groups & Validated groups \\
\midrule
    1 &        24 &    173 &               - &               - &                - \\
    2 &        33 &    218 &             264 &              75 &                8 \\
    3 &        38 &    289 &             868 &             113 &               11 \\
    4 &        37 &    232 &            1666 &              76 &               15 \\
    5 &        37 &    288 &            2016 &             101 &               22 \\
    6 &        38 &    276 &            1568 &             140 &               11 \\
    7 &        38 &    219 &             764 &             114 &               15 \\
    8 &        38 &    243 &             213 &              74 &               16 \\
    9 &        38 &    416 &              26 &              26 &               23 \\
\bottomrule
\end{tabular}
\end{center}
\caption{\label{tab:justiceScores2} Summary of the co-decisions of justices in groups of different size for the votations related to civil rights. For all sizes we report the number of justices active at least once and the number of votations in which groups of that size were observed. For all sizes larger than 1 we report the number of possible groups of co-voting justices (taking into account as a contrain the period of co-activity of justices), the number of observed groups and the number of groups validated in the SVH.}

\end{table}

\begin{table}
\small
\centering
\begin{center}
\begin{tabular}{rrrlll}
\toprule
 Size &  Justices &  Votes & Possible groups & Observed groups & Validated groups \\
\midrule
    1 &        28 &    234 &               - &               - &                - \\
    2 &        34 &    280 &             264 &             100 &                5 \\
    3 &        37 &    297 &             868 &             148 &                8 \\
    4 &        37 &    235 &            1666 &             135 &               12 \\
    5 &        38 &    295 &            2016 &             168 &               10 \\
    6 &        38 &    317 &            1568 &             188 &               19 \\
    7 &        38 &    278 &             764 &             156 &               19 \\
    8 &        38 &    302 &             213 &             109 &               28 \\
    9 &        38 &    525 &              26 &              26 &               23 \\
\bottomrule
\end{tabular}
\end{center}
\caption{\label{tab:justiceScores8} Summary of the co-decisions of justices in groups of different size for the votations related to economic issues. For all sizes we report the number of justices active at least once and the number of votations in which groups of that size were observed. For all sizes larger than 1 we report the number of possible groups of co-voting justices (taking into account as a contrain the period of co-activity of justices), the number of observed groups and the number of groups validated in the SVH.}

\end{table}

\section{High school students}

\begin{figure}[ht]
\centering
\includegraphics[width=0.95\textwidth]{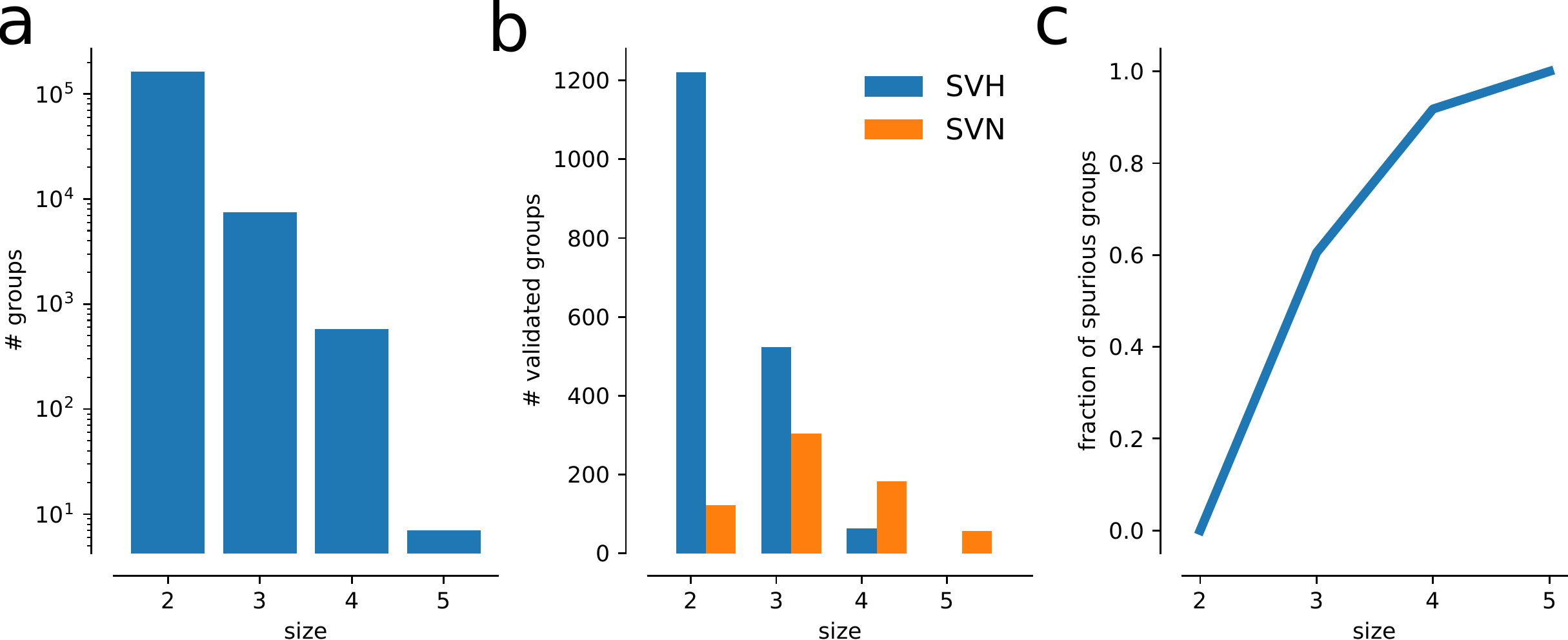}
\caption{\label{sociopatterns} \textbf{a)} Number of hyperlinks observed in the high school data as a function of size. \textbf{b)} Number of validated hyperlinks in the SVH (blue bars) and in the SVN (orange bars) as a function of size. \textbf{c)} Fraction of spurious hyperlinks (i.e. maximal pairwise cliques that do not correspond to observed hyperlinks) detected by the SVN as a function of size.
}
\end{figure}

In this section we present additional information on the analysis performed on the French high school dataset. Figure~\ref{sociopatterns}a plots the number of observed hyperlinks as a function of their size, showing the predominance of interactions at smaller sizes. The distribution of validated hyperlinks in the SVH (blue bars in Figure~\ref{sociopatterns}b) reproduces the same pattern of the unfiltered hypergraph, while the distribution of groups detected by the SVN (orange bars in Figure~\ref{sociopatterns}c) has a different shape. However, most of the groups detected by the SVN at higher sizes are spurious (i.e. they don't correspond to observed hyperlinks of that size in simultaneous interactions of students) as shown in Figure~\ref{sociopatterns}c.

To enrich the description of the hyperlinks validated in the SVH we compared both the SVH and the unfiltered hypergraph with the diaries data self reported by the student. Table~\ref{tab:diaries} reports the details of this comparison, showing how the SVH is more precise (i.e. it detect less spurious groups) but less accurate (i.e. it detects less true groups) of the unfiltered hypergraph.

\begin{figure}[ht]
\centering
\includegraphics[width=0.95\textwidth]{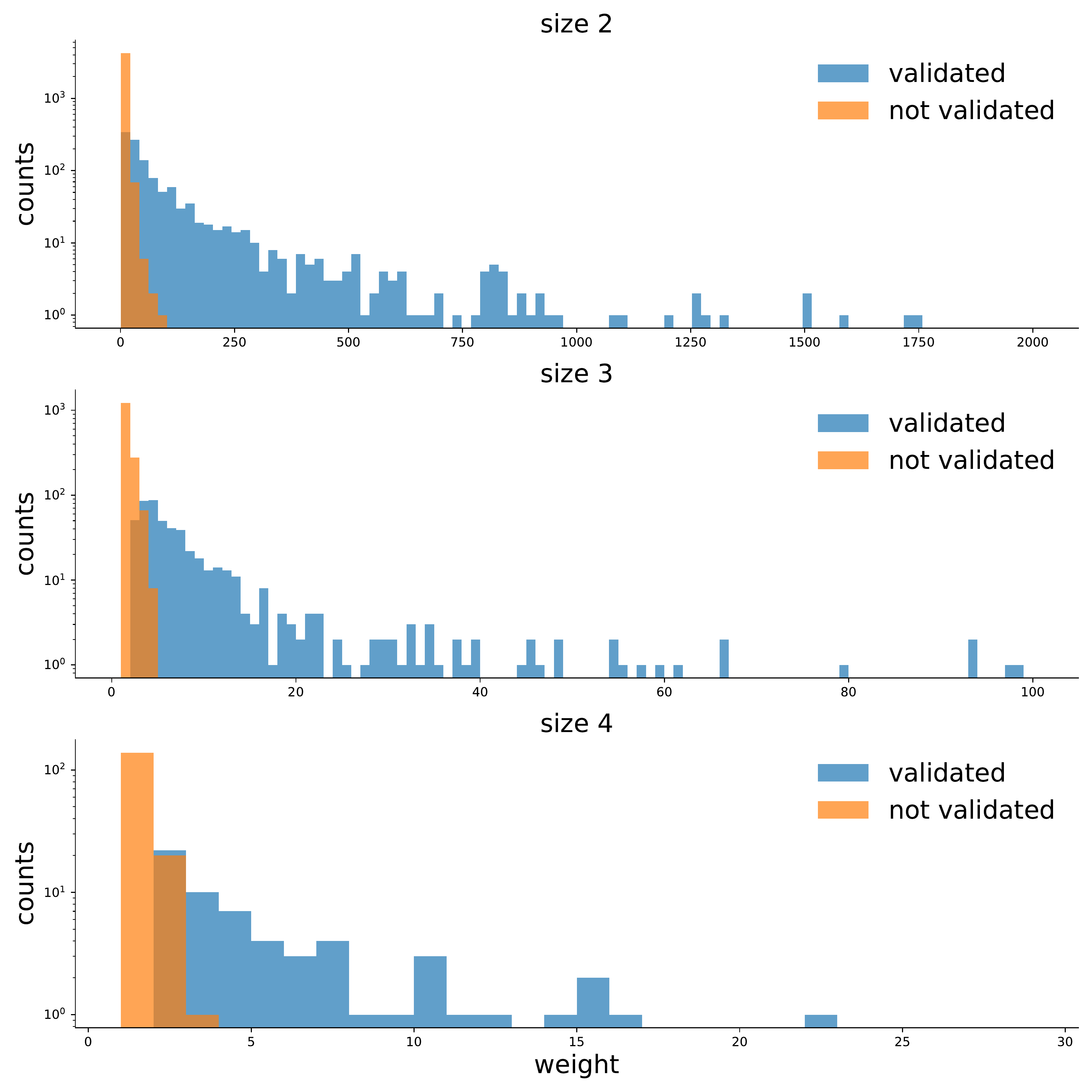}
\caption{\label{weights} Distribution of weights for the validated (blue bars) and the not validated (orange bars) hyperlinks at size 2 (top panel), size 3 (middle panel) and size 4 (bottom panel).
}
\end{figure}

At the end of the section in the main text we stress how the detection of validated hyperlinks in the SVH is more nuanced than a filtering procedure based on putting a threshold on weights. Figure~\ref{weights} shows the distributions of weights for the hyperlinks in the SVH and those that are not validated at different sizes. For all sizes there is an overlap in weights between the two groups. For these hyperlinks, the difference between being validated or not lies in the heterogeneous activity of its nodes. Indeed, for less active nodes hyperlinks with a small weight are more likely to be validated, while hyperlinks that involve more active nodes need a higher weight in order to be validated.

\begin{table*}
\small

\centering
\begin{center}
\begin{tabular}{r|r|r|r|r|r|r|r}
\toprule
size &  \# diaries &  \# SVH &  \# unfiltered &  SVH $\cap$ &  unfiltered $\cap$  & spurious &  spurious \\
~ &  ~ &  ~ &  ~ &  diaries &  diaries   & SVH & unfiltered \\
\midrule
    2 &        202 &    351 &          1322 &                78 &                      191  &    0.78 &                 0.86 \\
    3 &         90 &     61 &           265 &                14 &                       30  &  0.77 &               0.89 \\
    4 &         16 &      3 &            14 &                 0 &                        0  &    1.00 &                 1.00 \\
\bottomrule
\end{tabular}
\end{center}
\caption{Summary of the comparison between the diaries cliques and the hyperlinks in the SVH and the unfiltered hypergraph. For each hyperlink size we report the number of groups in the diaries data, in the SVH and in the unfiltered hypergraph, followed by the intersection with the diaries and the fraction of spurious groups. }
\label{tab:diaries}
\end{table*}

\section{Physics authors}
\begin{figure}[ht]
	\centering
	\includegraphics[width=1\textwidth]{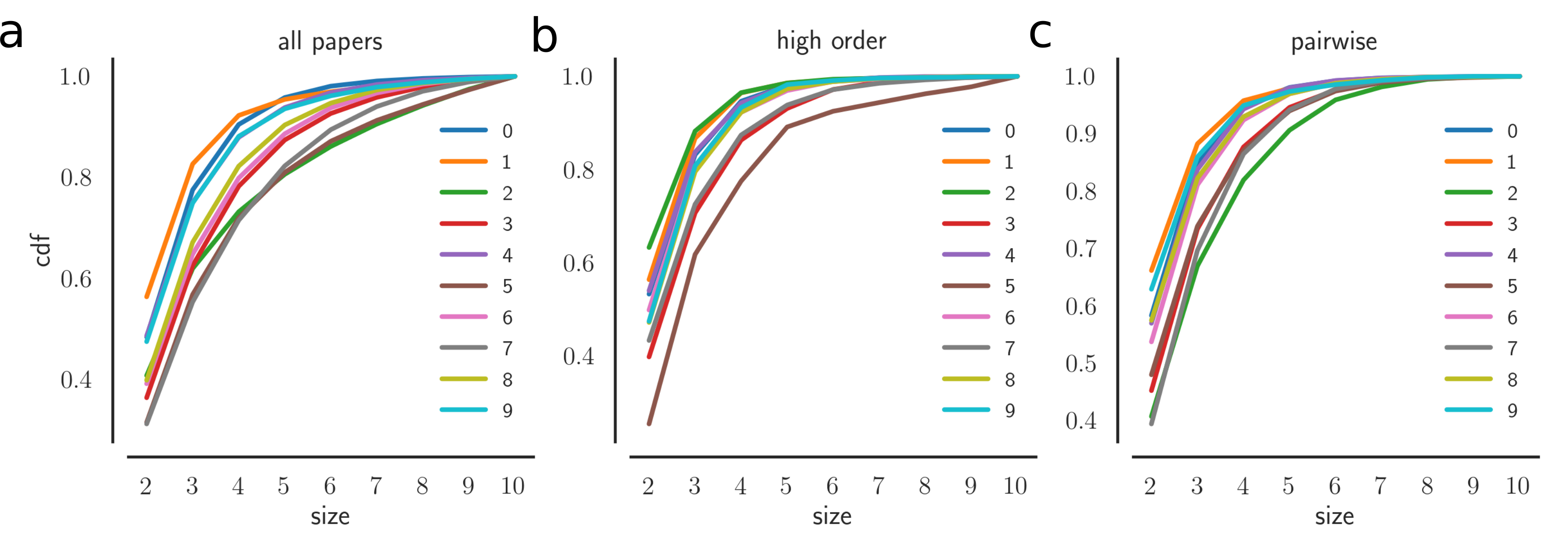}
	\caption{\label{fig:cdfs}\textbf{a)} Cumulative density function of papers as a function of team size for all ten PACS. \textbf{b)} Cumulative distribution function of hyperlinks in the SVH as a function of their size for all ten PACS. \textbf{c)} Cumulative distribution function of groups detected in the SVN as a function of their size for all ten PACS. For all panels, the legend reports the color code of the different PACS.}
	
\end{figure}

The APS dataset used in our analysis is the same of Ref: Battiston, Musciotto, Wang, Barabasi, Szell and Sinatra, Taking census of physics \textit{Nature Reviews Physics}, 1(1):89-97, 2019. In the main text we focus on three PACS, General Physics (PACS hierarchical integer number 0), Nuclear Physics (PACS number 2) and Physics of Gases and Plasma (PACS number 5). Figure~\ref{fig:cdfs} plots the cumulative density functions of papers (panel a), groups validated in the SVH (panel b) and in the SVN (panel c) for all ten PACS.

\begin{figure}[ht]
	\centering
	\includegraphics[width=1\textwidth]{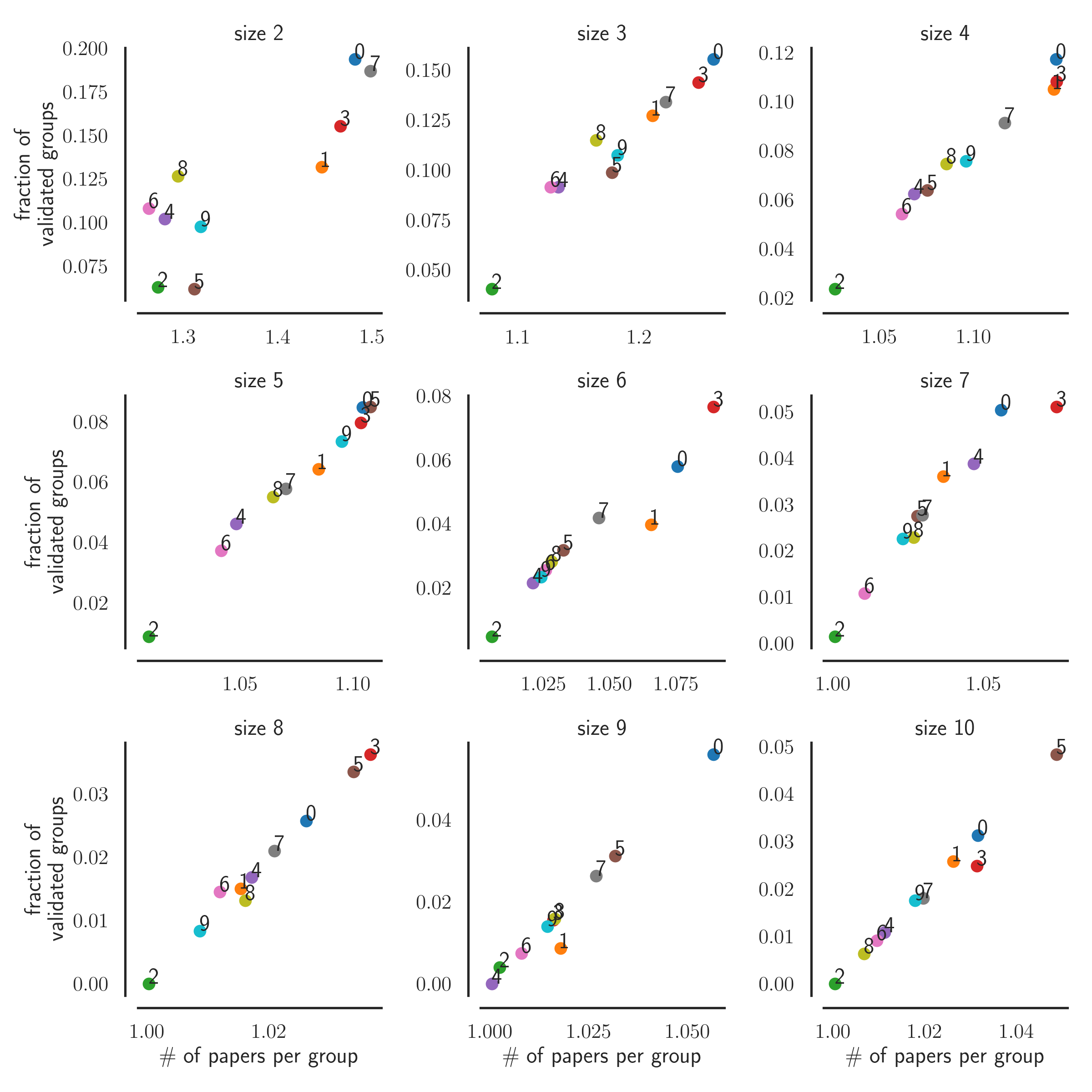}
\caption{\label{fig:corr_svh} Scatter plots of the average number of papers written by a group vs the fraction of validated hyperlinks in the SVH. Each dot represents a different PACS and each panel is related to a different group size. The color code for PACS is the same of Figure~\ref{fig:cdfs}. The number on each dot is the PACS identifier.}
	
\end{figure}

\begin{figure}[ht]
	\centering
	\includegraphics[width=1\textwidth]{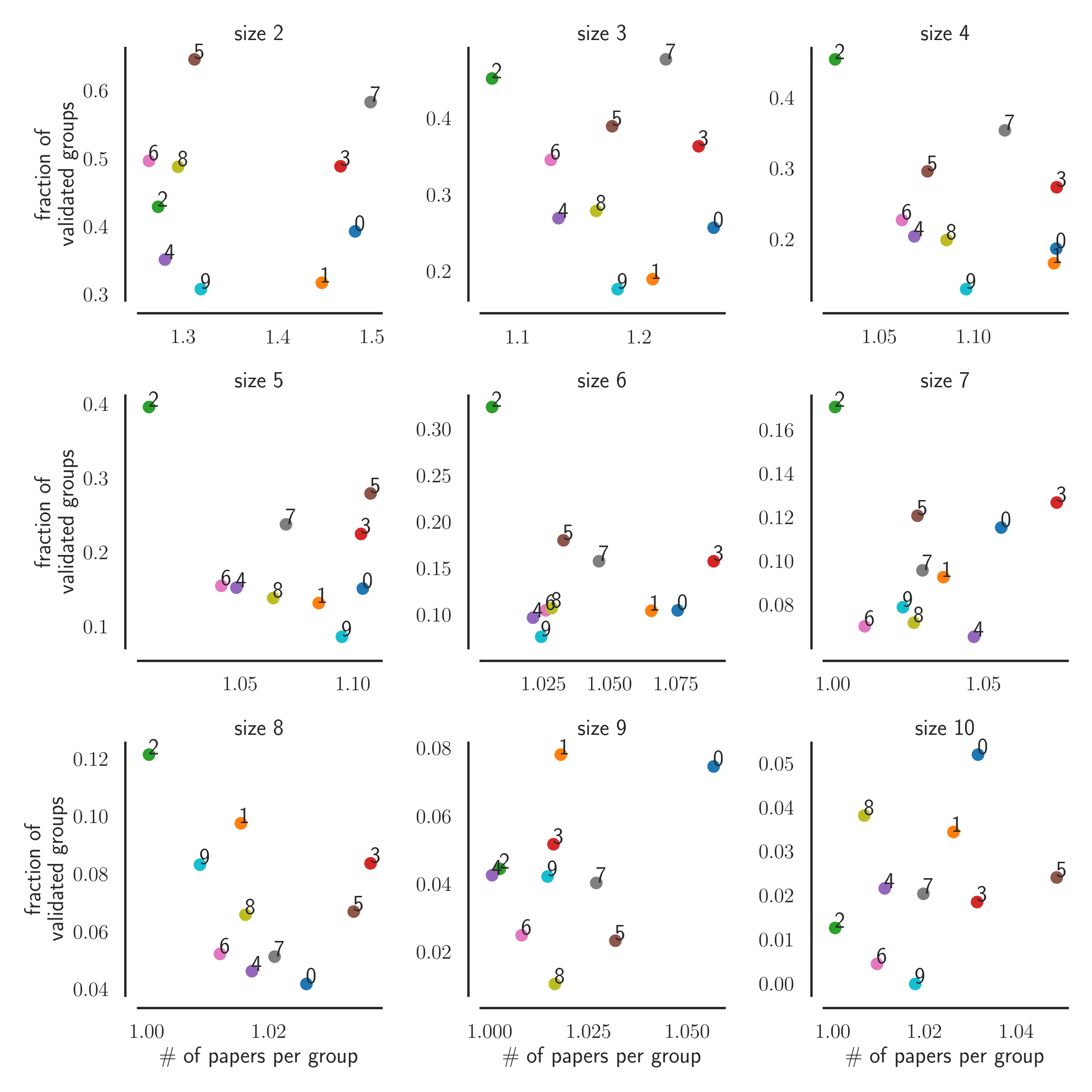}
	\caption{\label{fig:corr_svn}Scatter plots of the average number of papers written by a group vs the fraction of validated hyperlinks in the SVH. Each dot represents a different PACS and each panel is related to a different group size. The color code for PACS is the same of Figure~\ref{fig:cdfs}. The number on each dot is the PACS identifier.}
	
\end{figure}

In the main text we show that the correlation between the average number of papers written by research groups correlates with the fraction of groups validated with the SVH. Figure~\ref{fig:corr_svh} shows the scatter plots for all sizes of hyperlinks from 2 to 10. Figure~\ref{fig:corr_svn} shows the same plots for SVN. From the Figure we conclude that the two variables are not correlated.

